\documentclass[useAMS,usenatbib]{mn2e}

\usepackage{epsfig}

\begin{document}

\title{On the gravitational stability of  a galactic disc as a two-fluid system}

\author[M. Shadmehri and F. Khajenabi]{Mohsen Shadmehri\thanks{E-mail: m.shadmehri@gu.ac.ir; } and  Fazeleh Khajenabi\thanks{E-mail: f.khajenabi@gu.ac.ir;}\\
Department of Physics, Faculty of Science, Golestan University, Basij Square, Gorgan, Iran}

\maketitle

\date{Received ______________ / Accepted _________________ }

\begin{abstract}
Gravitational stability of a disc consisting of the gaseous and the stellar components are studied in the linear regime when the gaseous component is turbulent. A phenomenological approach is adopted to describe the turbulence, in which not only the effective surface density but the velocity dispersion of the gaseous component are both scale-dependent as power-law functions of the wavenumber of the perturbations. Also, the stellar component which has gravitational interaction with the gas is considered as a fluid. We calculate growth rate of the perturbations and in the most of the studied cases,  the stability of the disc highly depends on the existence of the stars and the  exponents of the functions for describing the turbulence. Our analysis suggests that the conventional gas and star threshold is not adequate for analyzing stability of the two-component discs when turbulence is considered.
\end{abstract}

\section{Introduction}
\label{sec:1}
 Various instability mechanisms may operate in an accretion disc depending on the physical properties of the disc itself. An imbalance of the heating and cooling rates in an thin accretion disc may lead to a disc which is thermally unstable \citep[e.g.,][]{LF74,pringle76,piran78}. Under some conditions, an accretion disc is hot enough to be fully or partially ionized which then implies to expect significant dynamical effects of the magnetic field. In fact, an instability related to the rotational profile and the magnetic field of an accretion disc is known to be the driven mechanism of generating turbulence inside the accretion discs \citep{balbus91}. In the most of the accreting systems, the mass of the disc is negligible in comparison to the mass of the central object and so, one may safely neglect the self-gravity of the disc. However, gaseous or stellar discs are subject to the gravitational instability if the Toomre parameter drops below unity \citep{toomre64}  and the cooling time scale is less than a few time of the dynamical time scale \citep{gam2001}.  Formation of structures like planets in protoplanetary discs \citep[e.g.,][]{boss98,rafikov2005,rafikov2007} or stars at the Galactic center \citep[e.g.,][]{tan2004,nayakshin2005,nayakshin2006,levin2007} are described based on the gravitational instability.

Analysis of the gravitational stability of a disc started by the  pioneer works of \citet{safr60} and \citet{toomre64}.  The onset
 of axisymmetric gravitational instability is determined by the following condition \citep{toomre64}
\begin{equation}\label{eq:Toomre}
Q=\frac{c_{\rm s} \kappa}{\pi G \Sigma} < Q_{0},
\end{equation}
where $Q_{0}=1$ and $c_{\rm s}$, $\kappa$ and $\Sigma$ are the sound speed, epicyclic frequency and the surface density of the disc, respectively. The non-dimensional Toomre parameter $Q$ has a vital role in the gravitational stability of an accretion disc. The non-axisymmetric perturbations  grow for somewhat higher $Q_{0}$.

However, the Toomre condition (\ref{eq:Toomre}) is not the only criteria for the fragmentation of a disc. In fact,  thermodynamics of the disc and its efficiency of cooling dictates another condition for the fragmentation of a disc as has been suggested in a seminal work by \citet{gam2001} and investigated via extensive numerical simulations \citep*[e.g.,][]{gammie2003,rice2005,clarke2007,cos2009}.

Although validity of the Toomre condition for the stability of the self-gravitating discs has also been confirmed by direct numerical simulations, in some of the astrophysical systems like galaxies one component description is not adequate to study their stability \citep[e.g.,][]{Yim2011,bournaud,bruce2011}. A typical galaxy consists of the gas and  the stars. Over recent two decades, many authors studied gravitational stability of such two-components discs \citep[e.g.,][]{jog84,romeo92,silk94,jog96}. The main outcome is the stellar component has a destabilizing role either a fluid approximation or a collisionless description is used to model the stellar component. Unfortunately, it is not possible to represent the stability condition of a two-component disc is a closed analytical form  like the one component case, though some authors introduced approximate analytical formulae for the onset of the instability in a disc \citep[e.g.,][]{silk94,romeo2011b}.

Recently, \citet*{romeo2010} (hereafter; RBA) studied possible effects of the turbulence on the gravitational stability of the accretion discs using a simple, but very illustrative model for the turbulence. They used scale-dependent relations  for the effective surface density and  the velocity dispersion of the gaseous component. Then, they could obtain conditions under which the self-gravitating disc becomes unstable depending on the properties of the turbulence. For a range of the input parameters, they showed the conventional Toomre parameter is modified  to correctly describe the stability of the disc when turbulence of the gas is considered.    However, such kind of approach for studying gravitational stability of the discs has already been applied by a few authors, but for some specific cases (e.g., Elmegreen 1996). In another relevant study, \citet*{bruce2011} investigated gravitational instabilities in two-component disc galaxies when the gas dissipates on the local crossing time. However, he did the linear analysis by considering the energy equation and prescribing a scale-dependent dissipation rate instead of  following a phenomenogical approach to describe the turbulence. The main finding is an increase to the conventional gas and star threshold by a factor of two or three because of the existence of the dissipation which removes the stabilization from turbulent pressure.

In this paper, we study gravitational stability of a two fluid disc where the turbulence of the gaseous component is considered similar to RBA. However, we will show that different regimes of the (in)stability according to RBA are significantly modified in the presence of the stars. One can also compare our results to \citet{jog84} to see how stability of a two fluid disc is changed when the turbulence of the gaseous component is considered. In the next section, our basic assumptions and the main dispersion relation are presented. Then, we will analyze stability of the system for a wide range of the input parameters and the results are compared to RBA.

\section{Linear Perturbations}
The basic equations of our model are identical to that introduced in \citet{jog84}, in particular regarding to the two-fluid description of the
system. Gaseous and the stellar components are considered as two separate fluids which can only interact gravitationally. Although we assume the stellar component is collisional (i.e., fluid description), it is also possible to consider the collisionless nature of the stellar component \citep{rafikov2001}. Both the gaseous and the stellar components are assumed to be isothermal characterized by the sound speed $\sigma_{\rm g}$ and the velocity dispersion $\sigma_{\rm s}$, respectively. The unperturbed surface densities  for the gas and the stellar fluid are denoted by $\Sigma_{\rm g0}$ and $\Sigma_{\rm s0}$. Also, the scale-heights of the gaseous component  and the   stellar mass distribution are given by $2h_{\rm g}$ and $2h_{\rm s}$.     But, we include turbulence of the gaseous component in a way that was introduced by RBA. In this approach, not only the initial density but also velocity dispersion of the gaseous component are both scale-dependent because of the existence of the turbulence:
\begin{equation}\label{eq:sigmaeff}
\Sigma_{\rm eff} = \Sigma_{\rm g0} \left ( \frac{k}{k_{0}}\right )^{-a},
\end{equation}
\begin{equation}\label{eq:sigmag}
\sigma_{\rm g} = \sigma_{\rm g0} \left ( \frac{k}{k_{0}} \right )^{-b},
\end{equation}
where $a$ and $b$ are input parameters which describe the nature of the turbulence. Density fluctuations has a power-law spectrum, $E_{\rho}(k) \propto k^{-r}$, which implies $\rho \propto k^{-(r-1)/2}$ \citep[see, e.g,][]{elmegreen2004,scalo}. Assuming $\Sigma_{\rm eff} \sim \rho h_{\rm g}$, one can easily show that $a=(1/2)(r-1)$.  Also, the spectrum of the velocity fluctuations is power-law, i.e. $E_{\rm v}(k) \propto k^{-s}$, and then $\sigma_{\rm g} \propto k^{-(s-1)/2}$. The above power-law relations are appropriate for the cold interstellar medium and the length $1/k_0$ is the fiducial scale at which Toomre parameter and other stability quantities are measured (RBA). Moreover,  the mass-size scaling relation implies $-2 \leq a \leq 1$ (RBA). However, observational and theoretical studies imply more constraints on the acceptable ranges for $a$ and $b$. HI observations show that a Kolmogorov spectrum can describe the density and the velocity fluctuations in these regions, i.e. $a\sim 1/3$ for the scales less than 10 kpc, and $b\sim 1/3$ for the scales less than 1 kpc \citep[e.g.,][]{lazar2000,Dutta2009}. But  in giant molecular clouds, the density and the velocity exponents are $a\sim 0$ and $b\sim 1/2$  \citep[e.g.,][]{Larson81,heyer}. On the other hand, high and low resolution simulations of supersonic turbulence imply the typical values of $(a,b)$ as $(1/2,1/2)$ and $(2/3,1/2)$, respectively (RBA).

\begin{figure}
\epsfig{figure=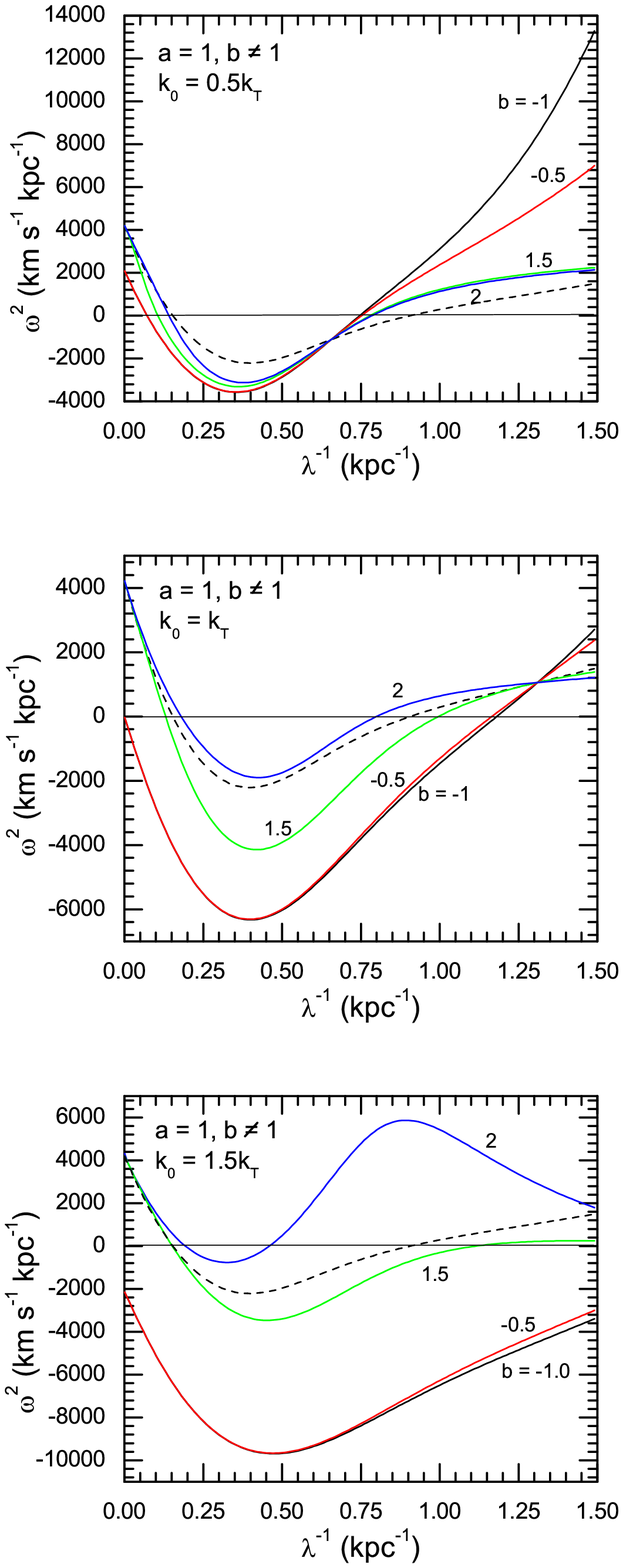,angle=0,scale=0.7}
\caption{This Figure shows $\omega^2$ versus $\lambda$ when $a=1$ and $b\neq 1$ and  $k_0=0.5 k_{\rm T}$ (top), $k_0 = k_{\rm T}$ (middle) and $k_{0} = 1.5 k_{\rm T}$ (bottom). The other input parameters are $\kappa = 65 $ ${\rm km} $ ${\rm s}^{-1} $ ${\rm kpc}^{-1}$, $\sigma_{\rm s}=34.7 $ ${\rm km}$  ${\rm s}^{-1}$, $\sigma_{\rm g0}=5 $ ${\rm km}$  ${\rm s}^{-1}$,  $\Sigma_{\rm g0}=19$ $M_{\odot}$ ${\rm pc}^{-2}$ and  $\Sigma_{\rm s0}=190$ $M_{\odot}$ ${\rm pc}^{-2}$. Dashed curve corresponds to a case without turbulence.}
\label{fig:f1}
\end{figure}

\begin{figure}
\epsfig{figure=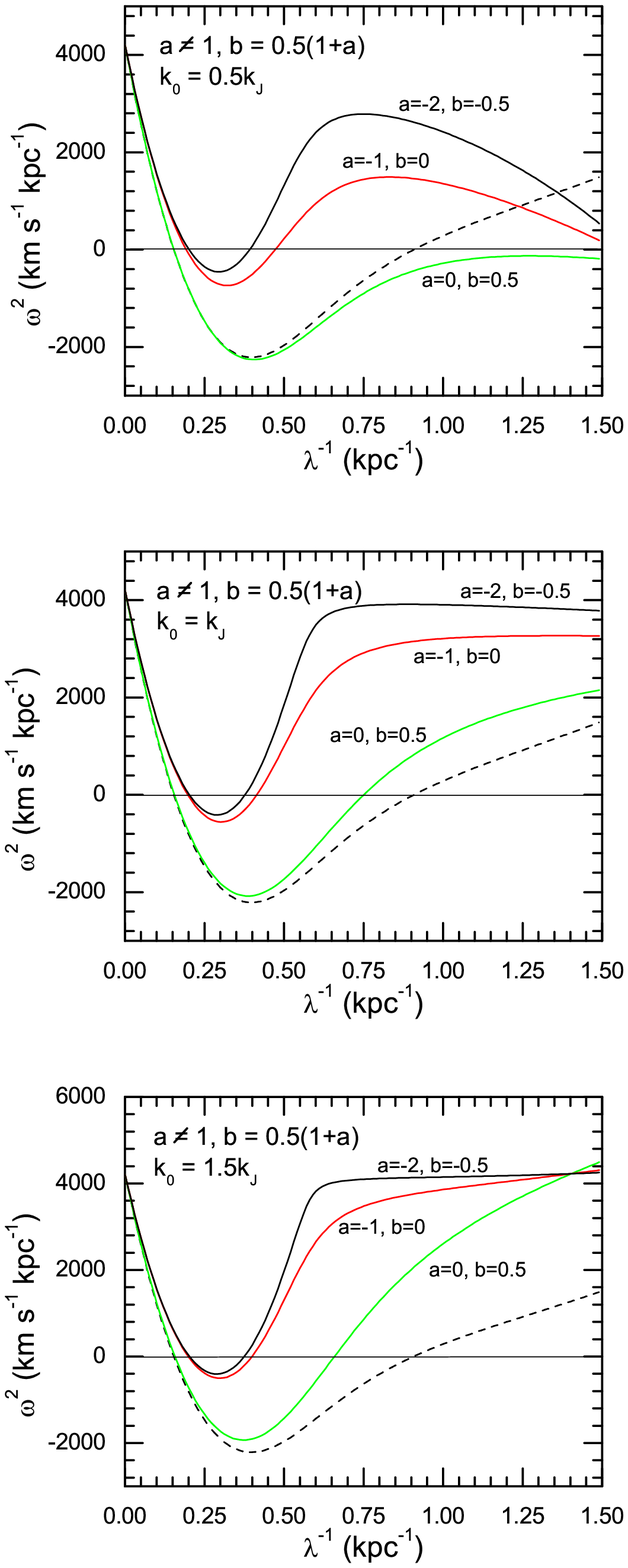,angle=0,scale=0.7}
\caption{The same as Figure \ref{fig:f1}, but for the case $a \neq 1$ and $b=0.5 (1+a)$.}
\label{fig:f2}
\end{figure}

\begin{figure}
\epsfig{figure=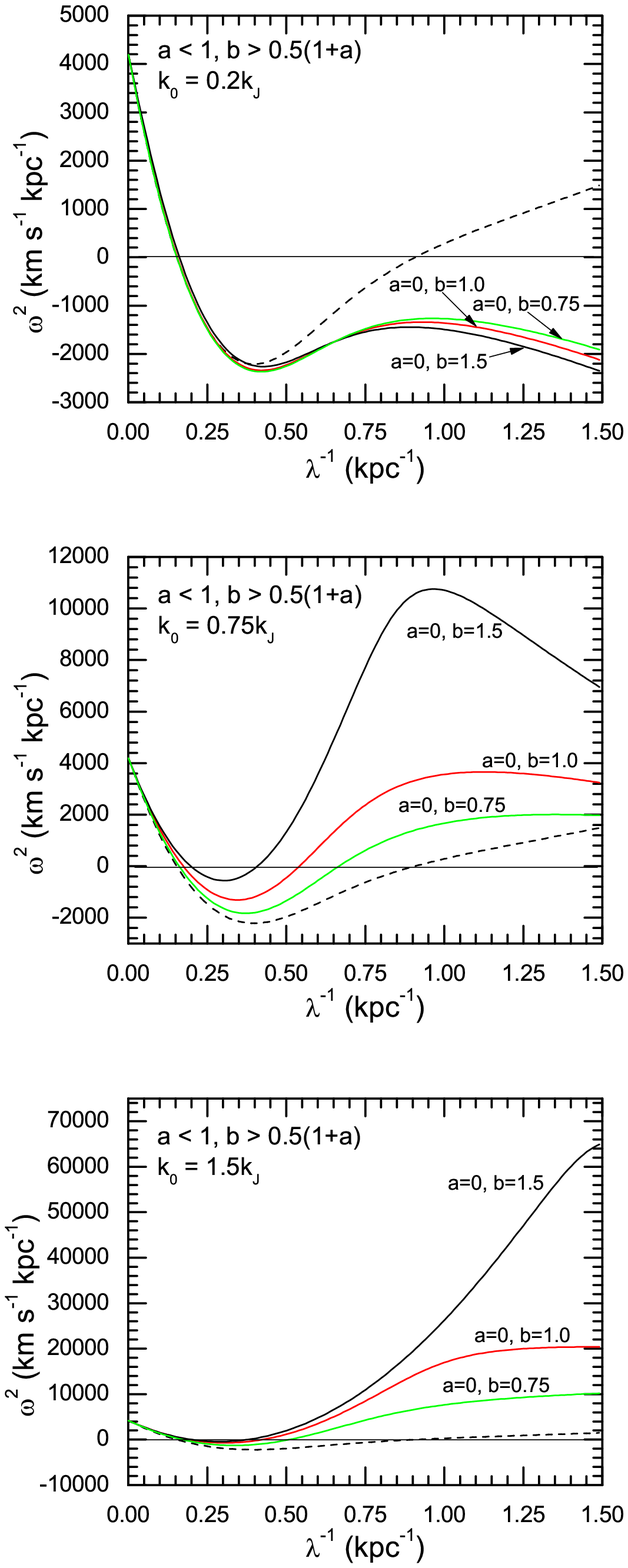,angle=0,scale=0.7}
\caption{The same as Figure \ref{fig:f1}, but for the case $a<1$ and $b>0.5 (1+a)$. Each curve is labeled by the corresponding values of $a=0$ and $b=0.75, 1, 1.5$.}
\label{fig:f3}
\end{figure}

\begin{figure}
\epsfig{figure=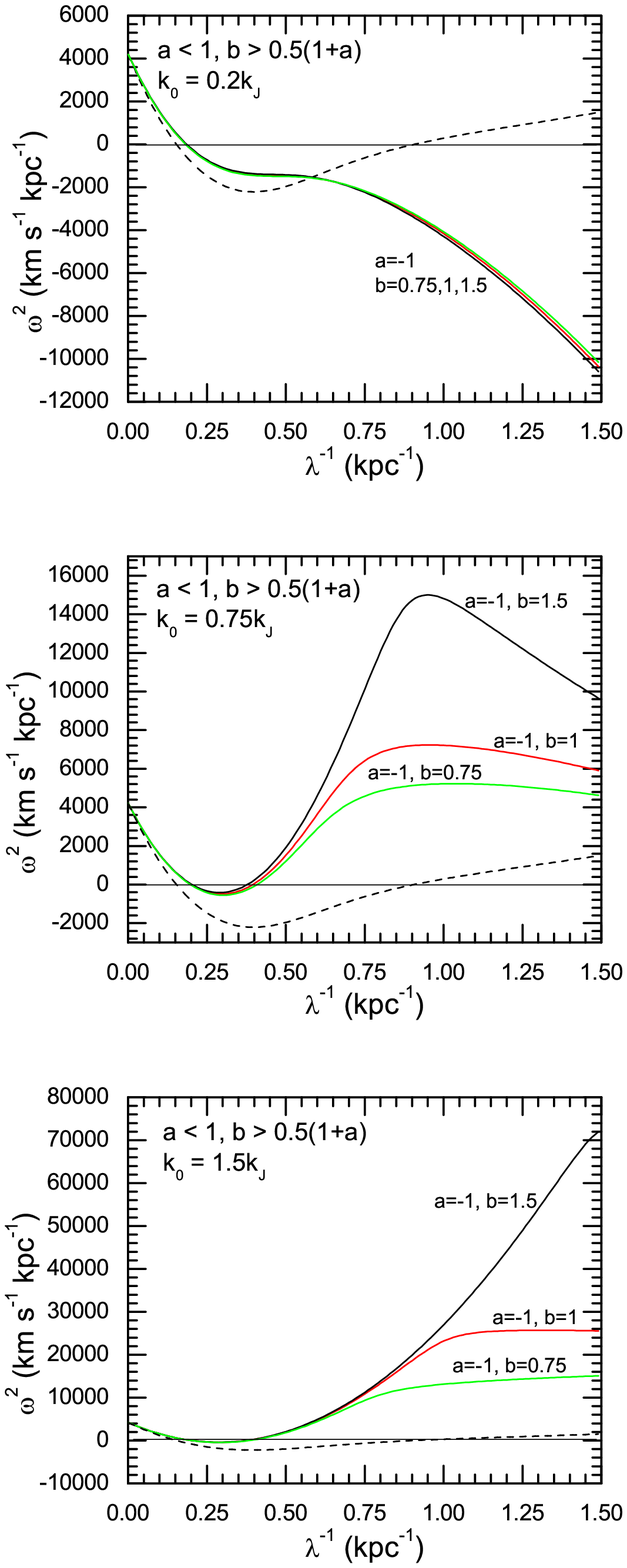,angle=0,scale=0.7}
\caption{The same as Figure \ref{fig:f1}, but for the case $a<1$ and $b>0.5 (1+a)$. Here, the input parameters are $a=-1$ and $b=0.75, 1, 1.5$}
\label{fig:f4}
\end{figure}

\begin{figure}
\epsfig{figure=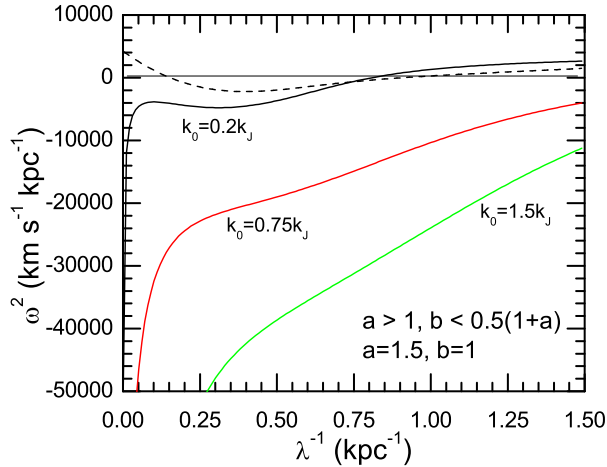,angle=0,scale=0.45}
\caption{The same as Figure \ref{fig:f1}, but for the case $a>1$ and $b<0.5 (1+a)$. Each curve is labeled by the corresponding value of $k_0$. The input parameters are $a=1.5$ and $b=1$.}
\label{fig:f5}
\end{figure}

If we assume a functional form as $\exp [i (kr+\omega t)]$ for the perturbed variables in the linearized hydrodynamical equations,  the dispersion relation is obtained as
\begin{equation}\label{eq:quad}
\omega^4 - \omega^2 (\alpha_{\rm s}+\alpha_{\rm g}) + (\alpha_{\rm s}\alpha_{\rm g} - \beta_{\rm s}\beta_{\rm g})=0,
\end{equation}
where
\begin{equation}
\alpha_{\rm s}=\kappa^2 + k^2 \sigma_{\rm s}^2 - 2\pi G k \Sigma_{\rm s0} \left[ 1-\exp(-kh_{\rm s})/k h_{\rm s} \right],
\end{equation}
\begin{equation}
\alpha_{\rm g}=\kappa^2 + k^2 \sigma_{\rm g}^2 - 2\pi G k \Sigma_{\rm eff} \left[ 1-\exp(-kh_{\rm g})/k h_{\rm g} \right] ,
\end{equation}
\begin{equation}
\beta_{\rm s}= 2\pi G k \Sigma_{\rm s0}  \left[ 1-\exp(-kh_{\rm s})/k h_{\rm s} \right],
\end{equation}
\begin{equation}
\beta_{\rm g}= 2\pi G k \Sigma_{\rm eff} \left[ 1-\exp(-kh_{\rm g})/k h_{\rm g} \right].
\end{equation}
We identify $\omega $ as the growth rate of the gas and star gravitational instability.
Since only the two-dimensional forcing is considered, a correction factor of $[1-\exp(-kh_{\rm s})]/(kh_{\rm s})$ appeared in the above equations \citep{toomre64}.  However, \citet{vander} introduced a correction factor of $(1+kh_{\rm s})^{-1}$ from a more detailed analysis. Note that the appropriate  correction factor  appears  for the gas as well.   Thus, if we set $a=b=0$, it means that the correction factors are considered for the stellar fluid and the non-turbulent gas component and the equations are reduced to JS. Equations (\ref{eq:quad}) is solved analytically as
\begin{equation}
\omega^2(k)= \frac{1}{2} \left [ (\alpha_{\rm s} + \alpha_{\rm g}) \pm \sqrt{(\alpha_{\rm s} + \alpha_{\rm g})^2 - 4 (\alpha_{\rm s} \alpha_{\rm g} -\beta_{\rm s} \beta_{\rm g})} \right ],
\end{equation}
where only the following root leads to the instability (JS),
\begin{equation}\label{eq:growth}
\omega^2(k)= \frac{1}{2} \left [ (\alpha_{\rm s} + \alpha_{\rm g}) - \sqrt{(\alpha_{\rm s} + \alpha_{\rm g})^2 - 4 (\alpha_{\rm s} \alpha_{\rm g} -\beta_{\rm s} \beta_{\rm g})} \right ].
\end{equation}
We note that if the stellar or the gaseous contributions to the (in)stability of the system are neglected, then growth rate of the unstable perturbations is determined based on the sign of $\alpha_{\rm s}$ or $\alpha_{\rm g}$. In other words, the stellar or the gaseous components are unstable when $\alpha_{\rm s}<0$ or $\alpha_{\rm g}<0$, respectively.

For a purely gaseous disc, RBA studied the instability regimes corresponding to $\alpha_{\rm g}<0$ for different values of $a$ and $b$. Now, we illustrate how this parameter study is modified when the stellar contribution is considered. Thus, we must determine under what circumstances equation (\ref{eq:growth}) gives a negative value, i.e. $\omega^2 <0 $. However, it is very unlikely to present the stability condition based on the growth rate equation (\ref{eq:growth}) in a closed analytical form. Instead, we adopt the parameters corresponding to the Galaxy as an illustrative example, but for different values of $a$ and $b$. For a gaseous disc, RBA showed that the stability of the disc is classified into seven categories depending on the values of $a$ and $b$. Now, we explore how these different regimes of the instability are modified in the presence of a stellar fluid.

\section{analysis}

JS adopted their input parameters for the Galaxy based on the observational data. We also consider a similar set of the input parameters, because it helps us to compare our results to the previous findings. In all plots, our input parameters correspond approximately to conditions at the distance $R=6$ ${\rm kpc}$ from  center of the Galaxy. Although the stars in the disc have a range of velocity dispersions, we characterize the stellar fluid by a single velocity dispersion $\sigma_{\rm s}=34.7 $ ${\rm km}$  ${\rm s}^{-1}$. The epicyclic frequency is $\kappa = 65 $ ${\rm km} $ ${\rm s}^{-1} $ ${\rm kpc}^{-1}$. The sound speed in the gas and the ratio of the gas density to the stellar density are $5$ ${\rm km}$ ${\rm s}^{-1}$ and $0.1$, respectively. Also, we assume $\Sigma_{\rm g0}=19$ $M_{\odot}$ ${\rm pc}^{-2}$ and  $\Sigma_{\rm s0}=190$ $M_{\odot}$ ${\rm pc}^{-2}$. We also assume the disc is in vertical equilibrium and the scale heights are assumed $2h_{\rm g}\approx 150 $ pc and $2h_{\rm s} \approx 180$ pc (JS).

Having the above input parameters, we can plot growth rate versus the wavenumber of the perturbations using equation (\ref{eq:growth}). Figure \ref{fig:f1} shows $\omega^2$ versus the wavelength of the perturbations, $\lambda$, when $a=1$ and $b\neq 1$. In this case, RBA showed that a gaseous disc is stable for all wavenumbers as long as $k_{0} \leq k_{\rm T} = \kappa^2 /(2\pi G \Sigma_{g0})$. Here, $k_{\rm T}$ is the Toomre wavenumber.  But we found this condition is not valid when the stellar fluid is considered. In this case, we note that for the gaseous component the self-gravity term is independent of the wavenumber, but for the stellar fluid it is not.  The Kolmogorov turbulence corresponds to $a=1$ and $b=1/3$ which lies in the explored regime of Figure \ref{fig:f1}. HI observations also suggest a Kolmogorov scaling for the velocity and the density fluctuations \citep[e.g.,][]{lazar2000,Dutta2009}.  The growth rate curves in Figure \ref{fig:f1} are drawn for three values of $k_0$, i.e. $k_{0}=0.5 k_{\rm T}$ (top plot), $k_{0}=k_{\rm T}$ (middle plot) and $k_{0}=1.5 k_{\rm T}$ (bottom plot). In all plots, dashed curves correspond to a case without turbulence, i.e. $a=b=0$. Top plot of Figure \ref{fig:f1} clearly shows the system is not stable for all the wavenumbers and always there is a range of $k$ where the disc is unstable, irrespective of the value of $b$. But when the critical wavenumber $k_0$ increases, more interesting cases emerge according to the middle and bottom plots of Figure \ref{fig:f1}. For example, RBA predicts that for $k_{0}=1.5 k_{\rm T}$ a gaseous disc is always unstable, but bottom plot of Figure \ref{fig:f1} shows a stability behavior when $b=2$. Also, the range of $k$ for which the disc is unstable becomes larger when $k_0$ increases.

In Figure \ref{fig:f2}, we explore another case: $a\neq 1$ and $b=0.5(1+a)$. So, both the pressure and the self-gravity terms of the gaseous component have the same $k-$dependence. Each curve is labeled by the corresponding values of $a$ and $b$ and again the dashed curves show growth rate of the perturbations when the turbulence of the gas is neglected.  Three different values of $k_0$ are adopted. For this case, RBA showed that  a gaseous disc is stable for all $k$ as log as $k_{0} \geq k_{\rm J}=(2\pi G \Sigma_{\rm g0})/ \sigma_{\rm g0}^2 $, where $k_{\rm J}$ is the conventional Jeans wavenumber. However, we see that this condition is modified at least for some of the input parameters, e.g. $a=0$ and $b=0.5$.

\begin{figure}
\epsfig{figure=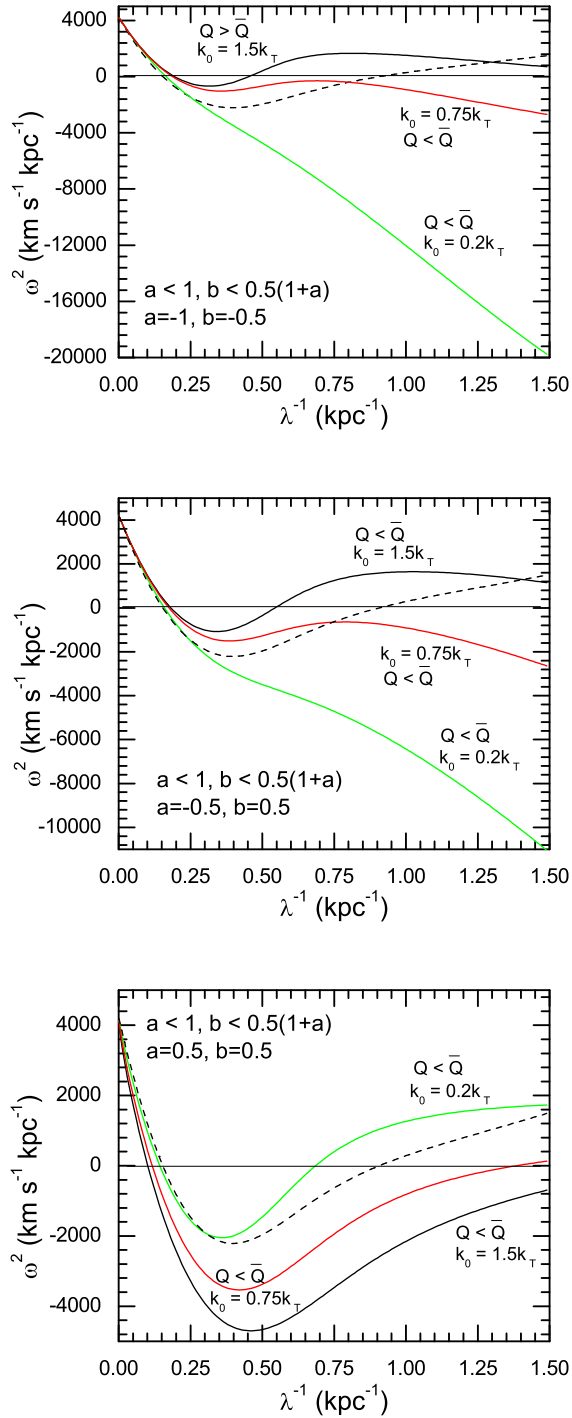,angle=0,scale=0.7}
\caption{The same as Figure \ref{fig:f1}, but for the case $a<1$ and $b<0.5 (1+a)$. As we see different values of $a$ and $b$ are considered.}
\label{fig:f6}
\end{figure}

Figures \ref{fig:f3} and \ref{fig:f4} show growth rate of the perturbations for the case $a<1$ and $b>0.5 (1+a)$. RBA did not explore this case in detail because the zero(s) of $\omega^2 (k)$ should be determined numerically, case by case. But they mentioned one component  gaseous disc is unstable at small scales, irrespective of the existence of the turbulence. Here, we also study this case for a limited range of the input parameters. For a particular set of the parameters $a$ and $b$ from this category, \citet{Elm96}  has also studied stability of a turbulent gaseous disc corresponding to the Larson-type scaling relations, i.e. $a=-1$ and $b=1/2$.  He found that the disc is always stable at large scales and unstable at small scales. In both Figures \ref{fig:f3} and \ref{fig:f4}, different values of $k_0$ are considered. We can see the disc becomes more stable with increasing $k_0$ so that when $k_0 = 1.5 k_J$ the system is stable for the whole  values of $a$ and $b$. But when $k_0 = 0.2 k_J $ and $a=0$, the system is unstable at two intervals of  the wavenumber. These intervals of the instability diminish with increasing $k_0$. Comparing  Figures \ref{fig:f3} and \ref{fig:f4}, we can say the system becomes more stable as the value of $a$ decreases. So, not only the results of RBA but Elmegreen's result are modified in the presence of a stellar fluid.

Figure \ref{fig:f5} shows growth rate of the perturbations for another considered case by RBA, i.e. $a>1$ and $b<0.5 (1+a)$. For this case, RBA showed that a turbulent gaseous disc is unstable at large scale. Our Figure \ref{fig:f5} confirms this behaviour for a turbulent gaseous disc in the presence of  the stellar fluid.

The case $a<1$ and $b<0.5 (1+a)$ is a Toomre-like situation as discussed by RBA. They found a stability threshold $\bar{Q}$ in terms of the input parameters and the turbulent gaseous disc is stable at all wavenumbers if $Q \geq \bar{Q}$.  Figure \ref{fig:f6} explores this case. Each curve is labeled by the relation between  the corresponding parameters $Q$ and ${\bar Q}$ based on RBA. For example, curves corresponding to $k_0 = 1.5 k_{\rm T}$ and $k_0 = 0.75 k_{\rm T}$ in the middle plot show stability in contrary to inequality $Q<{\bar Q}$. In other plots, the system becomes unstable for the limited ranges of the wavenumber of the perturbations. In all the plots, we also note that for fixed $\Sigma_{\rm s0}$ and $\Sigma_{\rm g0}$, varying $k_0$ has the effect of varying the effective surface density. Thus, the system has higher effective surface density and becomes more unstable at lower $k_0$ when $a<0$. But for positive values of $a$, we have less effective surface density and more stability at lower $k_0$.

\section{conclusions}

We studied gravitational stability of a disc consisting of two components (i.e., gaseous and stellar fluids) using a perturbation analysis. We also considered the turbulent nature of the gaseous fluid using  a phenomenological description which is supported by  the observations and the numerical simulations. Although stability of a two-fluid disc  without turbulence of the gas \citep*[e.g.,][]{jog84,jog96,silk94,rafikov2001} or one component  turbulent gaseous disc \citep*{romeo2010} have already been studied analytically, but our analysis is a first step towards considering not only the two-fluid nature of such systems but the turbulence of the gas. We found that turbulence of gas highly affects the stability of the two-component stellar and fluid discs. According to our analysis, not only there is not a closed analytical relation for onset of the instability in a two-component disc, but the conventional gas and star threshold does not correctly address the stability of such systems in the presence of turbulence of the gas. We also explored situations without the thickness correction for the gas. In fact, this correction factor has a stabilizing effect, in particular at high wavenumber $k$. For the perturbations much larger than the thickness of the disc, the thickness correction becomes unity. But as the wavenumber increases so that for the scales larger than the thickness of the disc, the correction factor introduces an additional inverse $k$ dependence to the effective surface density. This weakens the self-gravitational effect significantly. In the presence of the turbulence of the gas, we also found this stabilizing effect.

In dwarf galaxies or the far-outer regions of the galaxy discs, the conventional stability criteria suggests a high level of stability \citep[e.g.,][]{van,hunter}. Although \citet*{bruce2011} showed that  such regions are more unstable when  turbulence of the gas is considered, our analysis shows that the stability of such systems highly depends on the nature of the turbulence (i.e., values of $a$ and $b$) in the presence of stars. However, detailed numerical simulations of  the galaxy discs (including the turbulence of the gas component) are needed to confirm our linear stability analysis.

\section*{Acknowledgments}
We are grateful to the anonymous referee whose detailed and careful comments helped to improve the quality of this paper.

\bibliographystyle{mn2e}
\bibliography{reference}

\end{document}